\documentclass[12pt]{iopart}
\usepackage{graphicx}

\begin{document}

\def\beq{\begin{equation}}
\def\eeq{\end{equation}}
\def\ber{\begin{eqnarray}}
\def\eer{\end{eqnarray}}
\def\l{\Lambda}
\def\lsim{\
\lower-1.5pt\vbox{\hbox{\rlap{$<$}\lower5.3pt\vbox{\hbox{$\sim$}}}}\ }
\def\gsim{\
\lower-1.5pt\vbox{\hbox{\rlap{$>$}\lower5.3pt\vbox{\hbox{$\sim$}}}}\ }
\def\apj{{Astroph.\@ J.\ }}
\def\mn{{Mon.\@ Not.@ Roy.\@ Ast.\@ Soc.\ }}
\def\asta{{Astron.\@ Astrophys.\ }}
\def\aj{{Astron.\@ J.\ }}
\def\prl{{Phys.\@ Rev.\@ Lett.\ }}
\def\prd{{Phys.\@ Rev.\@ D\ }}
\def\nucp{{Nucl.\@ Phys.\ }}
\def\nat{{Nature\ }}
\def\plb {{Phys.\@ Lett.\@ B\ }}
\def \jetpl {JETP Lett.\ }
\def\ie {{i.e.}}
\def\n {\noindent}
\def\etal{{\it et al.}}
\def\m{{\rm m}}
\def\b{{\rm b}}
\def\d3h{\lower-1pt\hbox{$\stackrel{\ldots}{H}$}}
\def \ie {i.e.}
\def \lleq {\lower0.9ex\hbox{ $\buildrel < \over \sim$} ~}
\def \ggeq {\lower0.9ex\hbox{ $\buildrel > \over \sim$} ~}

\title[Quantum effects, soft singularities and the fate of the universe]
{Quantum effects, soft singularities and the fate of the universe in a
braneworld cosmology}

\author{Petr Tretyakov\dag,\ Aleksey Toporensky\dag,\
Yuri Shtanov\ddag\ and Varun Sahni\S}
\address{$\dag$\ Sternberg Astronomical Institute, Moscow State University, Universitetsky
Prospekt, 13, Moscow 119899, Russia\\
$\ddag$\ Bogolyubov Institute for Theoretical Physics, Kiev 03143, Ukraine\\
$\S$\ Inter-University Centre for Astronomy and Astrophysics, Post Bag 4,
Ganeshkhind, Pune 411~007, India}




\begin{abstract}
We examine a class of braneworld models in which the expanding universe
encounters a ``quiescent'' future singularity. At a quiescent singularity, the
energy density and pressure of the cosmic fluid as well as the Hubble parameter
remain finite while all derivatives of the Hubble parameter diverge (i.e.,
${\dot H}$, ${\ddot H}$, etc.\@ $\to \infty$). Since the Kretschmann invariant
diverges ($R_{iklm}R^{iklm} \to \infty$) at the singularity, one expects
quantum effects to play an important role as the quiescent singularity is
approached. We explore the effects of vacuum polarization due to massless
conformally coupled fields near the singularity and show that these can either
cause the universe to recollapse or, else, lead to a softer singularity at
which $H$, ${\dot H}$, and ${\ddot H}$ remain finite while $\d3h$ and higher
derivatives of the Hubble parameter diverge. An important aspect of the
quiescent singularity is that it is encountered in regions of low density,
which has obvious implications for a universe consisting of a cosmic web of
high and low density regions\,---\,superclusters and voids. In addition to
vacuum polarization, the effects of quantum particle production of
non-conformal fields are also likely to be important. A preliminary examination
shows that intense particle production can lead to an accelerating universe
whose Hubble parameter shows oscillations about a constant value.

\end{abstract}

\pacs{04.20.Dw, 04.50.+h, 04.62.+v, 98.80.-k}

\maketitle

\section{Introduction}

Braneworld models give rise to interesting new physical effects
\cite{maartens,sahni04}. The well-known Dvali--Gabadadze--Porrati (DGP) model,
for instance, can lead to an accelerating universe without the presence of
either a cosmological constant or some other form of dark energy \cite{DGP}.
Generalizations of the DGP model can result in a phantom-like acceleration of
the universe at late times \cite{SS,lue}. This class of models can also lead to
new cosmological behaviour at {\em intermediate redshifts}: the loitering
\cite{loiter} and mimicry \cite{mimicry} scenarios provide examples of
cosmologies which are close to LCDM at late times but can show significant
departure from LCDM-like expansion at $z \ggeq {\rm few}$. In both cases, the
age of the high-redshift universe turns out to be larger than in LCDM, while
the redshift of reionization is lower. Whether the universe has properties
which are easier to explain within the braneworld context is an intriguing
possibility demanding further exploration.  In this paper, we examine yet
another property of braneworld models which does not have parallels in general
relativity: the possibility that the universe may encounter a {\em quiescent
future singularity\/} as it expands \cite{SS1}.

It is well known that the rate of expansion of the universe and its ultimate
fate depend on the system of equations governing evolution (general relativity,
scalar-tensor theory, braneworld theory etc.\@) as well as on the form of
matter and, particularly, on its equation of state. In general relativity (GR),
if one assumes that matter satisfies the strong energy condition (SEC) $\rho +
3p \geq 0$, then, within a Friedmann--Robertson--Walker (FRW) setting, the
evolution of the universe is strongly dependent upon the spatial curvature: a
spatially closed universe turns around and collapses whereas open and flat
cosmologies continue to expand forever. The situation becomes more complicated
(and interesting\,!) if one of the following assumptions is made: (i)~the
expansion of the universe is not governed by GR, (ii)~matter can violate the
SEC and even the weak energy condition (WEC) $\rho + p \geq 0$. In the latter
case, if $w = p/\rho < -1$, then the expanding universe can encounter a ``big
rip'' future singularity at which the density, pressure, and Hubble parameter
diverge. In the former case, if the equations of motion have been derived from
a braneworld action \cite{SS}, then the expanding universe can encounter a
different kind of (quiescent) future singularity, at which the density,
pressure and Hubble parameter {\em remain finite\/}, but derivatives of the
Hubble parameter, including ${\dot H}$, diverge as the singularity is
approached \cite{SS1}. The occurrence of this singularity is related to the
fact that the equations of motion are no longer quasi-linear (as they are, for
instance, in GR) but include terms which are non-linear in the highest
derivative. Examples of such singularities can be found in models other than
the braneworld model; for instance, \cite{kamen04} refer to singularities in
which the deceleration parameter tends to infinity as the ``Big Brake'' while,
in \cite{barrow04}, they are called ``sudden'' singularities (see also
\cite{topor}). The geometrical reason of a quiescent singularity in braneworld
models is connected with the fact that the brane embedding in the bulk becomes
singular at some point (see \cite{SS1} for details).  Since quiescent
singularities can occur both in the past and in the future, they might provide
an interesting alternative to the more conventional ``big bang''/``big crunch''
singularities of general relativity. An important distinction between the
quiescent and the sudden singularity is the following: The existence of the
quiescent singularity does not require matter with unusual properties, hence,
both density and pressure remain finite near this singularity. For the sudden
singularity, on the other hand, the pressure diverges as the singularity is
approached, implying the presence of matter with exotic properties.

In this paper, we examine the issue of how quantum effects might influence a
braneworld which encounters a quiescent singularity during expansion. It is
well known that quantum effects come into play when the space-time curvature
becomes large, as happens, for instance, in the vicinity of a black hole or
near the Big-Bang and Big-Crunch singularities of general relativity \cite{BD}.
Since $R_{iklm}R^{iklm} \to \infty$ as one approaches a quiescent (sudden)
singularity, one might expect quantum effects to become important in this case
too (see, for instance, \cite{SS1,nojiri1}). As we demonstrate in this paper,
quantum corrections to the equations of motion at the semi-classical level
result in several important changes in the evolution of the universe: due to
the (local) effects of vacuum polarization, (i)~the quiescent singularity
changes its form and becomes a much weaker ``soft'' singularity, at which $H$
and ${\dot H}$ remain finite but $\d3h \to \infty$; (ii)~vacuum polarization
effects can also cause a spatially flat universe to turn around and collapse.
Both (i) and (ii) demonstrate that the incorporation of quantum effects into
the braneworld equations of motion can radically alter the future of the
universe and lead to behaviour which differs significantly from
general-relativistic cosmology. Since the quiescent singularity (and the
associated quantum effects) arise as the density drops below a threshold value,
it follows that those regions which are significantly underdense (voids) may be
the first to encounter the quiescent singularity. We also briefly discuss
whether particle production effects could be significant as the universe
approaches a quiescent future singularity. An interesting possibility that may
arise in this case is that the universe expands in a regime in which the Hubble
parameter vacillates about the de Sitter value.\,\footnote[4]{It should be
noted that braneworld models which approach the first quiescent singularity in
the future (and have Friedmann-like behaviour in the past) appear to be in some
tension with recent observational data \cite{rule-out}. However, the
observational status of the oscillatory model discussed in Sec.~\ref{steady}
remains to be studied.}

\section{Equations of motion}\label{equations}

We consider the simplest generic braneworld model with action of the form
\begin{equation} \label{action}
\fl S =  M^3 \left[\int_{\rm bulk} \left( {\cal R} - 2 \Lambda_{\rm b} \right)
- 2 \int_{\rm brane} K \right] + \int_{\rm brane} \left( m^2 R - 2 \sigma
\right) + \int_{\rm brane} L \left( h_{ab}, \phi \right) \, .
\end{equation}
Here, ${\cal R}$ is the scalar curvature of the metric $g_{ab}$ in the
five-dimensional bulk, and $R$ is the scalar curvature of the induced metric
$h_{ab} = g_{ab} - n_a n_b$ on the brane, where $n^a$ is the vector field of
the inner unit normal to the brane, which is assumed to be a boundary of the
bulk space, and the notation and conventions of \cite{Wald} are used. The
quantity $K = h^{ab} K_{ab}$ is the trace of the symmetric tensor of extrinsic
curvature $K_{ab} = h^c{}_a \nabla_c n_b$ of the brane. The symbol $L (h_{ab},
\phi)$ denotes the Lagrangian density of the four-dimensional matter fields
$\phi$ whose dynamics is restricted to the brane so that they interact only
with the induced metric $h_{ab}$. All integrations over the bulk and brane are
taken with the corresponding natural volume elements. The symbols $M$ and $m$
denote the five-dimensional and four-dimensional Planck masses, respectively,
$\Lambda_{\rm b}$ is the bulk cosmological constant, and $\sigma$ is the brane
tension.

Action (\ref{action}) leads to the Einstein equation with cosmological constant
in the bulk:
\begin{equation} \label{bulk}
{\cal G}_{ab} + \Lambda_{\rm b} g_{ab} = 0 \, ,
\end{equation}
while the field equation on the brane is
\begin{equation} \label{brane}
m^2 G_{ab} + \sigma h_{ab} = T_{ab} + M^3 \left(K_{ab} - h_{ab} K \right) \, ,
\end{equation}
where $T_{ab}$ is the stress--energy tensor of matter on the brane stemming
from the last term in action (\ref{action}), i.e.,
\begin{equation}
T_{ab} = {1 \over \sqrt{h}} {\delta \displaystyle \int_{\rm brane} L \left(
h_{ab}, \phi \right) \over \delta h^{ab}} \, .
\end{equation}

The five-dimensional bulk, satisfying Eq.~(\ref{bulk}), is described by the
metric
\begin{equation}
ds_{\rm bulk}^2 = - f (r) d \tau ^2 + \frac{dr^2}{f (r)} + r^2 d\Omega_\kappa^2
\, , \quad f (r) = \kappa - \frac{\Lambda_{\rm b}}{6}\, r^2 - {C \over r^2} \,
.
\end{equation}
Here, $\kappa = 0, \pm 1$ is the sign of the spatial curvature of the brane, $d
\Omega_\kappa^2$ denotes the metric of the maximally symmetric
three-dimensional Euclidean space with constant curvature corresponding to
$\kappa$, and the constant $C$, if it is nonzero, corresponds to the presence
of a black hole in the bulk. The trajectory of the brane in the bulk is given
by $r = a(\tau)$, and then the brane is made the boundary by discarding either
the region $r > a(\tau)$ or the region $r < a(\tau)$ from the bulk, resulting
in two possible cosmological branches (see Eq.~(\ref{embed}) below).

The cosmological evolution on the brane that follows from Eqs.~(\ref{bulk}) and
(\ref{brane}) can be encoded in a single equation \cite{SS,Shtanov02}
\begin{equation}\label{main}
\left(H^2 + {\kappa \over a^2} - {\rho + \sigma \over 3 m^2} \right)^2 = {4
\over \ell^2} \left( H^2 + {\kappa \over a^2} - {\Lambda_{\rm b} \over 6} - {C
\over a^4} \right) \, ,
\end{equation}
where $H \equiv \dot a/a$ is the Hubble parameter, and $\rho$ is the matter
energy density on the brane. Here and below, the overdot derivative is taken
with respect to the cosmological time $t$ on the brane, which is connected with
the bulk time $\tau$ through the obvious relation
\begin{equation}
{ dt \over d \tau} = \sqrt{f (a) - { \left( d a / d \tau \right)^2 \over f (a)
}} \, .
\end{equation}
The term containing the constant $C$ describes the so-called ``dark
radiation.''  The length scale $\ell$ is defined as
\begin{equation} \label{ell}
\ell = {2 m^2 \over M^3} \, .
\end{equation}

In what follows, we consider a spatially flat universe ($\kappa = 0$) without
dark radiation ($C = 0$).  Then Eq.~(\ref{main}) takes the form
\begin{equation}\label{main1}
\left(H^2 - {\rho + \sigma \over 3 m^2} \right)^2 = {4 \over \ell^2} \left( H^2
- {\Lambda_{\rm b} \over 6} \right) \, .
\end{equation}
This equation can be solved with respect to the total energy density on the
brane $\rho_{\rm tot} \equiv \rho + \sigma$:
\begin{equation}\label{embed}
{\rho + \sigma \over 3 m^2} = H^2 \pm {2 \over \ell} \sqrt{H^2 - {\Lambda_{\rm
b} \over 6} } \, .
\end{equation}
The ``$\pm$'' signs in the solution correspond to two branches defined by the
two possible ways of bounding the Schwarzschild--(anti)-de~Sitter bulk space by
the brane, as described above \cite{CH,Deffayet}.  Discarding the region $r >
a(\tau)$ (the region $r < a(\tau)$) from the bulk corresponds to the ``$+$''
(``$-$'') sign in (\ref{embed}).

Alternatively, Eq.~(\ref{main1})  can be solved with respect to $H^2$ with the
result \cite{SS,CH,Shtanov}
\begin{equation} \label{solution}
H^2 = {\rho + \sigma \over 3 m^2} + {2 \over \ell^2} \left[1 \pm \sqrt{1 +
\ell^2 \left({\rho + \sigma \over 3 m^2} - {\Lambda_{\rm b} \over 6} \right)}
\right] \, .
\end{equation}
Models with the lower (``$-$'') sign in this equation were called Brane\,1, and
models with the upper (``$+$'') sign were called Brane\,2 in \cite{SS}, and we
refer to them in this way throughout this paper.

\section{Classical dynamics of the Brane}\label{classical}

\begin{figure}
\begin{center}
\includegraphics[width=0.45\textwidth]{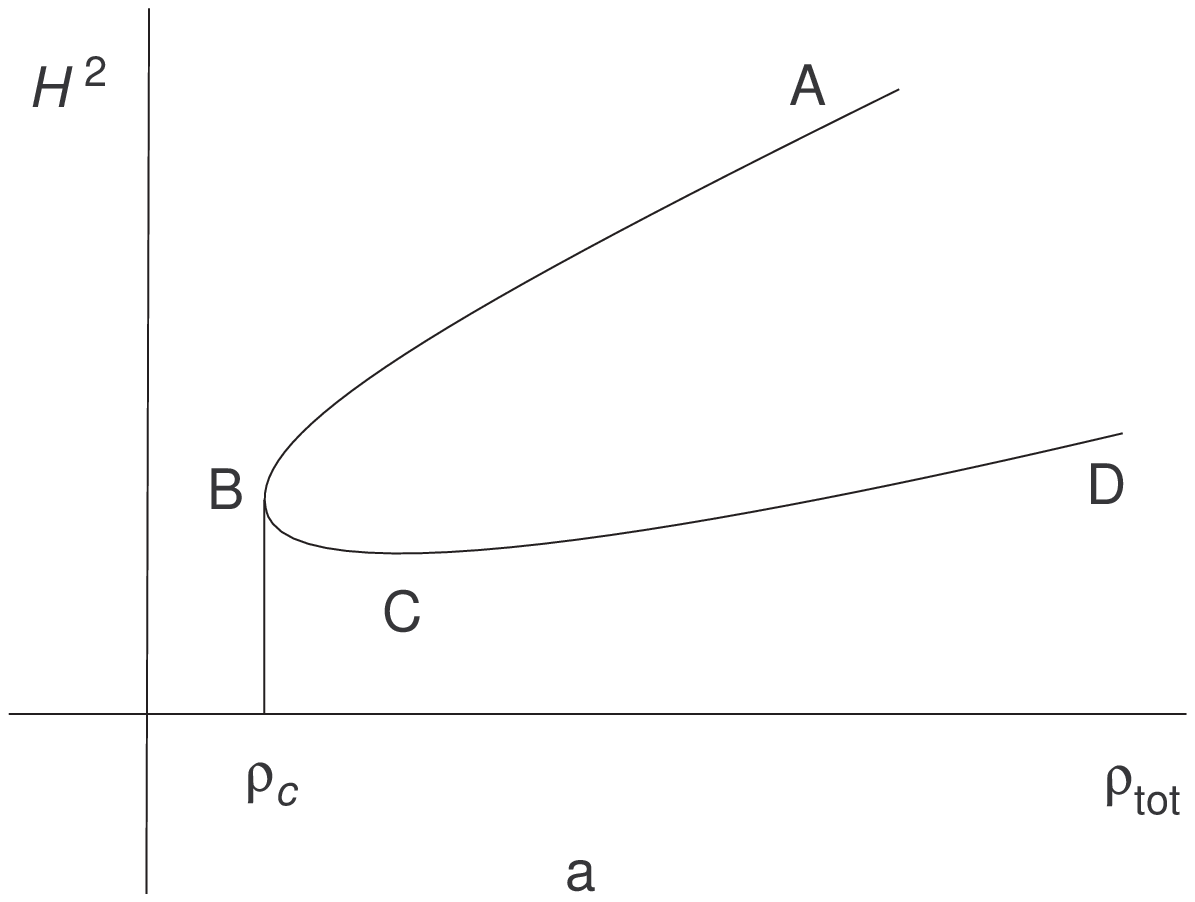} \hspace{1cm}
\includegraphics[width=0.45\textwidth]{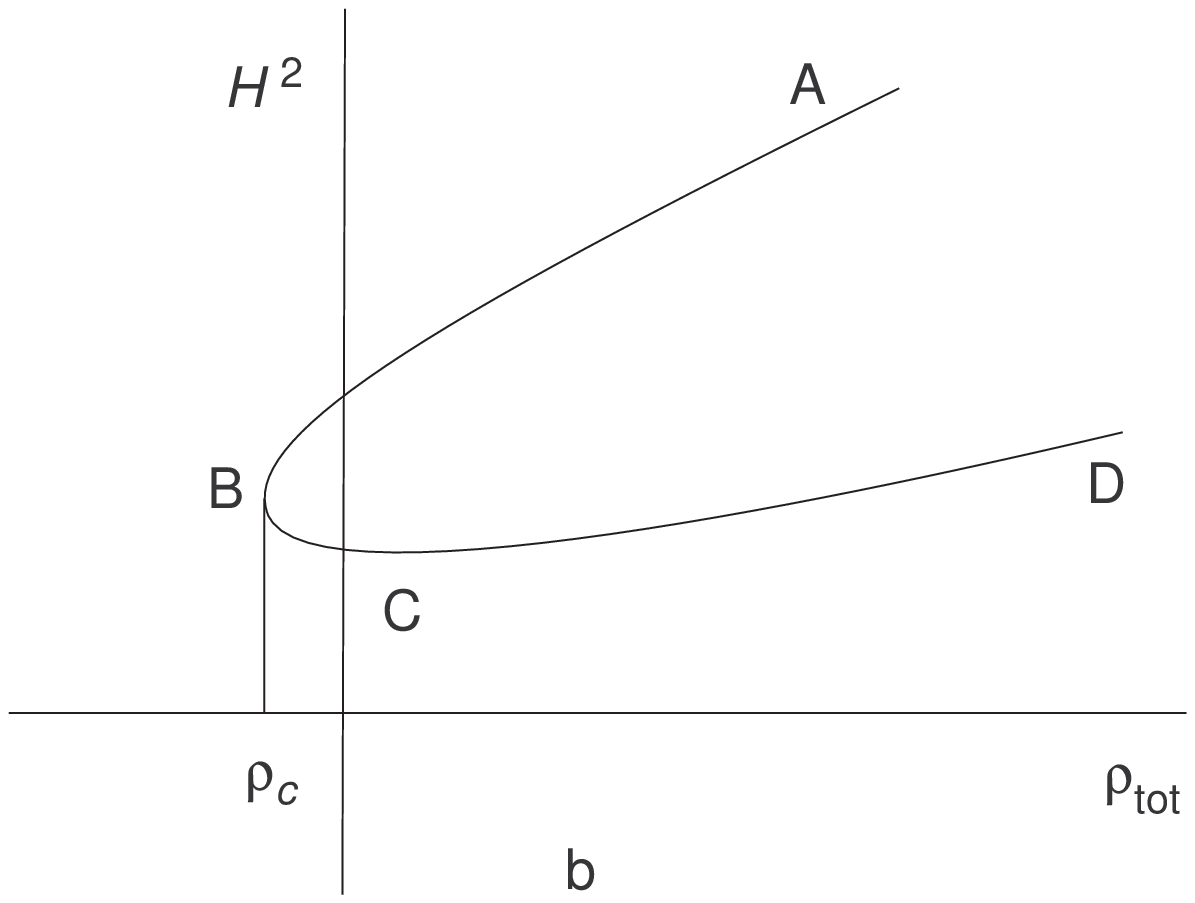}
\end{center}
\caption{\label{pos}Plot of relation (\ref{main1}) in the case $\Lambda_\b
> 0$. Case (a) corresponds to $\rho_c > 0$, and case (b) corresponds to $\rho_c
< 0$, where $\rho_c$ is given by Eq.~(\ref{rhoc}).}
\end{figure}

Before we study quantum corrections to brane equations of motion, we describe
possible classical dynamical regimes. Classical dynamics depends significantly
on the sign of the bulk cosmological constant $\Lambda_{\rm b}$. In the ensuing
discussion, we shall examine separately all three cases, namely $\Lambda_{\rm
b}>0$, $\Lambda_{\rm b}=0$, and $\Lambda_{\rm b}<0$.

\begin{itemize}
\item The case $\Lambda_{\rm b}>0$ is shown in Fig.~\ref{pos}. The graph of
(\ref{main1}) in the $\left( H^2, \rho_{\rm tot} \right)$ plane in
Fig.~\ref{pos} illustrates that in an expanding universe the matter density
$\rho$ decreases (except for a ``phantom matter'' which we do not consider in
the present paper), and the point in the plane $(H^2,\rho_{\rm tot})$ moves
from right to left in Fig.~\ref{pos}.

A striking feature of Fig.~\ref{pos} is that the value of the Hubble parameter
in the braneworld can {\em never drop to zero}. In other words, the Friedmann
asymptote $H \to 0$ is absent in our case. The upper and lower branches in
Fig.~\ref{pos} describe the two complementary braneworld models: branches AB
and DB are associated with Brane\,2 and Brane\,1 of \cite{SS}, respectively,
while branches AC and DC correspond to the lower and upper signs in
(\ref{embed}), respectively, and describe the two branches with different
embedding in the bulk. It should be noted that, in many important cases, the
behaviour of the braneworld does not have any parallel in conventional
Friedmannian dynamics (by this we mean standard GR in a FRW universe). For
instance, the BC part of the evolutionary track corresponds to ``phantom-like''
cosmology with $\dot H>0$, even though matter on the brane never violates the
weak energy condition.

We would like to draw the reader's attention to the fact that, for a given
$\rho_{\rm tot}$, a solution $H^2(\rho_{\rm tot})$ exists if and only
if\,\footnote{This corresponds to the quantity under the square root in
(\ref{solution}) being non-negative.}
\begin{equation}
\rho_{\rm tot} \ge \rho_c \equiv 3 m^2 \left(\Lambda_{\rm b}/6 -
\ell^{-2}\right)  \label{rhoc}
\end{equation}
(see Fig.~\ref{pos}). This leads to two distinct possibilities for the
late-time cosmological evolution of the braneworld. (i)~If $\sigma>\rho_c$,
then nothing prevents the matter density from diluting to $\rho \to 0$ at late
times. In this case, the braneworld approaches a De~Sitter-like future
attractor at which $H \to {\rm constant}$. (ii)~In the opposite case, when
$\sigma<\rho_c$, the braneworld dynamics is very different.  In this case,
Eq.~(\ref{rhoc}) can be rewritten as $\rho \geq \rho_c-\sigma$, which implies
that the density of matter can only drop to $\rho_c - \sigma$ and no further\,!
Indeed, at $\rho \to \rho_c-\sigma$, the universe experiences a ``quiescent''
singularity, at which the density $\rho$ and the Hubble parameter $H$ remain
finite, while derivatives of the $H$, including ${\dot H}$, ${\ddot H}$ etc.\@,
diverge \cite{SS1}. (For $\Lambda_{\rm b} < 6 \ell^{-2}$, we have $\rho_c < 0$,
and, in order for the quiescent future singularity to exist, the brane tension
$\sigma$ must be negative (see Fig.~\ref{pos}b).)

\item The case $\Lambda_{\rm b}=0$ is shown in Fig~\ref{zero}. The Minkowski
bulk ($\Lambda_{\rm b}=0$) leads to two generic future attractors, namely, the
De~Sitter-like Brane\,2 \cite{DGP} and the Friedmann-like braneworld evolving
to the point C (with $H \to 0$ and $\rho \to 0$). However, a sufficiently large
negative brane tension ($\sigma < -3 m^2 / \ell^{2}$) can also trigger the
formation of a quiescent future singularity at the point B.

\begin{figure}
\begin{center}
\includegraphics[width=0.5\textwidth]{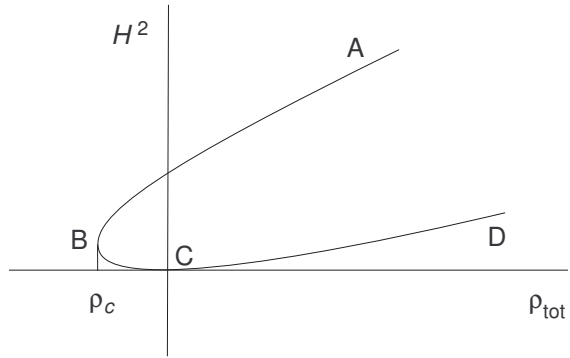}
\end{center}
\caption{\label{zero}Plot of relation (\ref{main1}) in the case $\Lambda_\b =
0$. The point C dividing the branches with different embedding corresponding to
different signs in (\ref{embed}) is at the origin of coordinates.}
\end{figure}

\item The case $\Lambda_{\rm b}<0$ is shown in Fig.~\ref{neg}. A very
interesting situation arises in this case since the Hubble parameter can reach
zero within a finite interval of time with $\dot H<0$ subsequently. This
situation describes the recollapse of the universe.

\begin{figure}
\begin{center}
\includegraphics[width=0.45\textwidth]{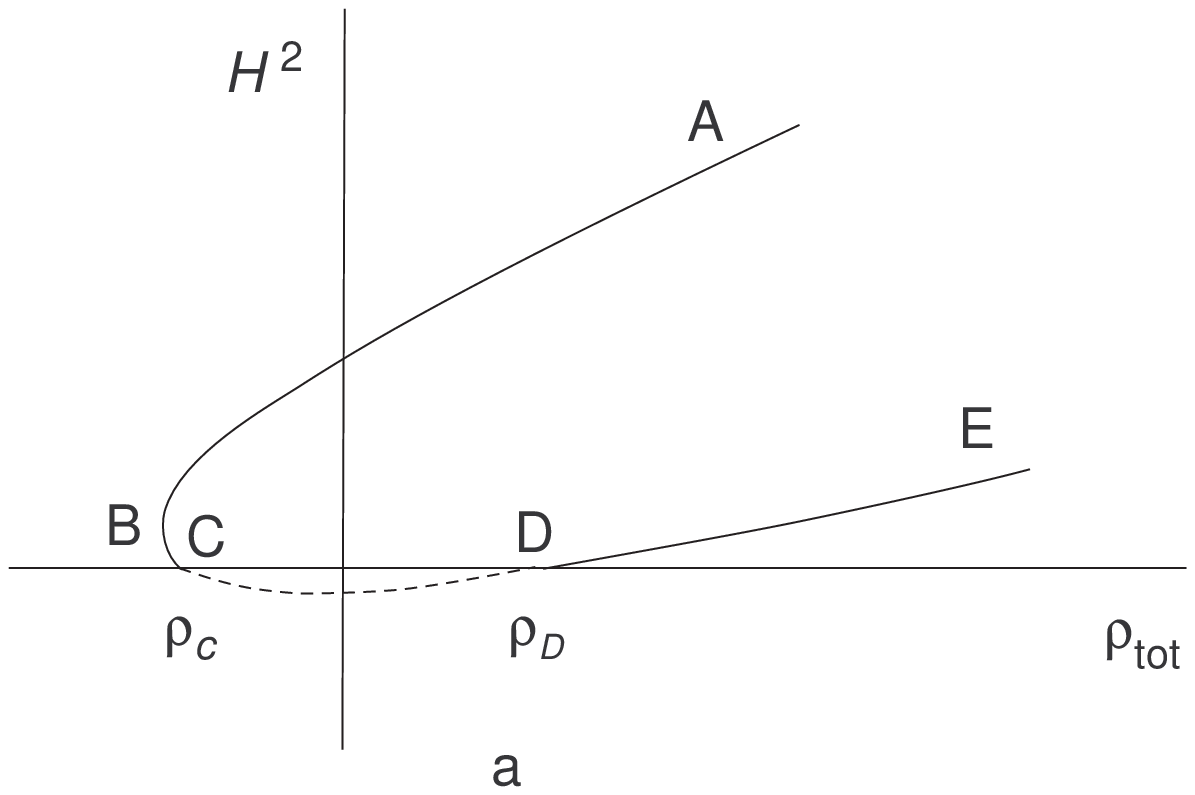} \hspace{1cm}
\includegraphics[width=0.45\textwidth]{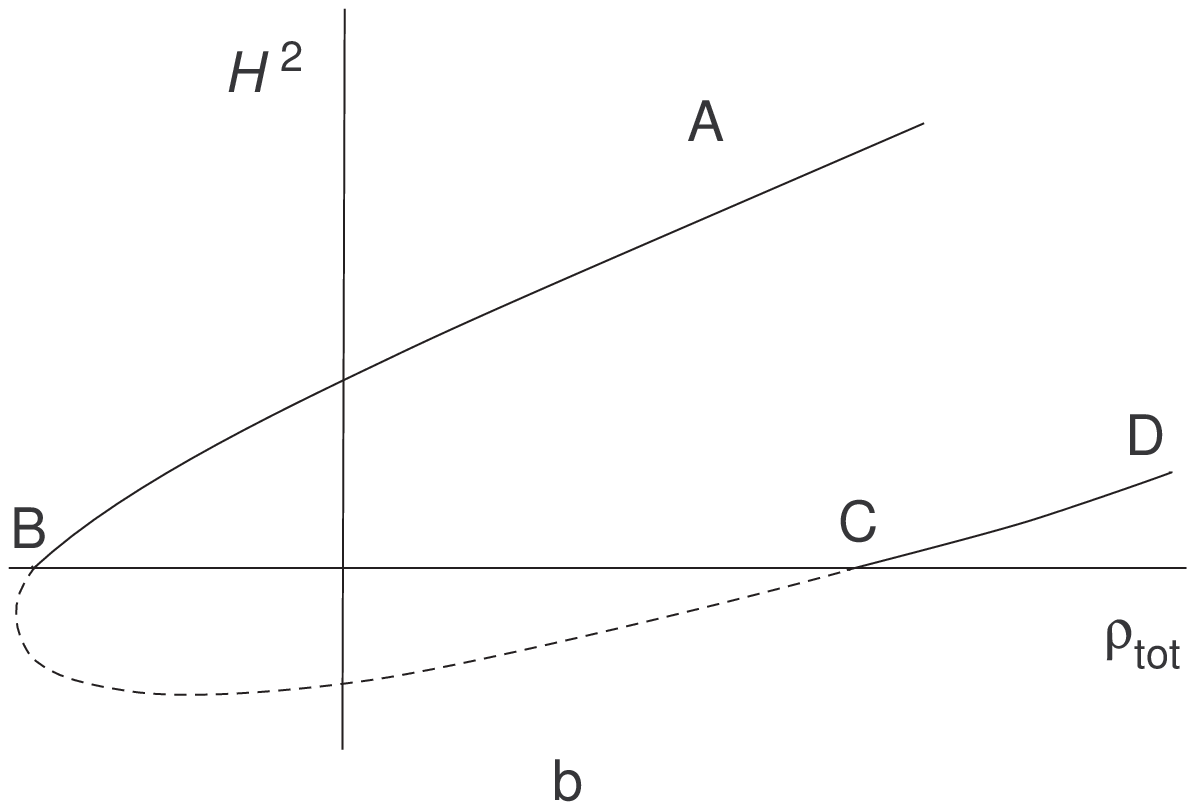}
\end{center}
\caption{\label{neg}Plot of relation (\ref{main1}) in the case $\Lambda_\b <
0$. Case (a) corresponds to the condition $|\Lambda_{\rm b}| < 6 \ell^{-2}$,
and case (b) corresponds to the condition $|\Lambda_{\rm b}| > 6 \ell^{-2}$.
The dashed part corresponds to the forbidden nonphysical branch.}
\end{figure}

It is worth stressing that the recollapse of the braneworld is a consequence of
the modified expansion law (\ref{solution}) and is not due to the presence of a
spatial curvature term\,---\,the recollapsing braneworld is spatially flat\,!

It is easy to show that Brane\,1 (lower branch) invariably recollapses provided
the brane tension is less then $\rho_D = M^3 \sqrt{ - 3 \Lambda_\b / 2}\,$,
which is the energy density at the point D in Fig.~\ref{neg}a. The critical
case $\sigma=\rho_D$ leads to a Friedmann late-time dynamics \cite{MMT}. The
Brane\,2 case (the upper branch AB), however, leads to two different
possibilities. Furthermore, this difference is important if we allow for the
existence of negative brane tensions. For instance, in Fig.~\ref{neg}a, a
Brane\,2 evolves along the AB part and encounters a quiescent singularity at
the point B. This requires negative tension $\sigma < 3 m^2 (\Lambda_{\rm
b}/6-\ell^{-2})$ \cite{SS1}. The part BC corresponds to a hypothetical universe
which is born at the singularity B, expands to the point C, then contracts and
ends up in the quiescent singularity at B.  The BC track is characterized by
the negative derivative of $\rho_{\rm tot}$ as a function of $H^2$. Simple
calculations lead to the condition $|\Lambda_{\rm b}| < 6 \ell^{-2}$ for the
existence of this branch. For $|\Lambda_{\rm b}| > 6 \ell^{-2}$, the situation
is presented in Fig.~\ref{neg}b. In this case, the Brane\,2 universe also
recollapses (provided the brane tension is negative and sufficiently large by
absolute value).

\end{itemize}

\section{Quantum corrections to the equations of motion}\label{quantum}

It is well known that quantum effects, including vacuum polarization and
particle production, generically occur within regions of strong space-time
curvature such as in the vicinity of black holes and near the cosmological
Big-Bang singularity. As we have seen earlier, the braneworld\,---\,in addition
to having the usual Big-Bang and Big-Crunch singularities\,---\,also possesses
quiescent singularities, which an observer can encounter while the universe
{\em is still expanding\/}. This singularity is specific to the braneworld,
since the density of matter, its pressure, and the Hubble parameter all freeze
to constant values, whereas ${\dot H}$, ${\ddot H}$ etc.\@ diverge as the
singularity is approached. Since this particular singularity can also develop
in an accelerating universe akin to ours \cite{SS1}, it is of considerable
interest to ask whether the nature of the quiescent singularity will in any way
be affected once quantum effects are incorporated into the treatment.

In general, quantum effects in curved space-time can arise on account of the
vacuum polarization as well as particle production. It is well known that the
latter is absent for conformally invariant fields (which we shall consider in
this section) and that, in this case, quantum corrections to the equations of
motion are fully described by the renormalized vacuum energy--momentum tensor
which has the form \cite{BD}
\begin{equation}
\left\langle T_{ab} \right\rangle = k_1 H_{ab}^{(4)} + k_2 H_{ab}^{(3)} + k_3
H_{ab}^{(1)} \, ,
\end{equation}
where
\begin{eqnarray}
H_{ab}^{(1)} &\equiv& 2 D_a D_b R - 2 h_{ab} D^c D_c R + \frac12 h_{ab} R - 2
R R_{ab} \, , \label{h1} \\
H_{ab}^{(3)} &\equiv& {}- R^c{}_a R_{cb} + \frac23 R R_{ab} + \frac12 h_{ab}
R^{cd} R_{cd} - \frac14 h_{ab} R^2 \, , \label{h2}
\end{eqnarray}
and $H_{ab}^{(4)}$ is a local non-geometric tensor which depends upon the
choice of the vacuum state and has vanishing trace $\left( H_a^{(4)a} = 0
\right)$; $k_1, k_2, k_3$ depend upon the spin weights of the different
fields contributing to the vacuum polarization.

Equations (\ref{h1}) and (\ref{h2}) lead to the following vacuum expectation
value for the energy density:
\begin{equation}
\rho_q \equiv \left\langle T_{00} \right\rangle
= k_2 H^4 + k_3 \left( 2 \ddot H H + 6\dot H H^2 - \dot H^2 \right) \, .
\label{rhoq}
\end{equation}
In order to assess the effects of the vacuum polarization on the dynamics of
the braneworld, one must add $\rho_q$ to the matter density in (\ref{main1}),
(\ref{solution}) or (\ref{embed}) so that $\rho \to \rho+\rho_q$ in those
equations. An important consequence of this operation is that the form of the
equation of motion changes dramatically\,---\,the original algebraic equation
changes to a differential equation\,! The dynamical equation (\ref{embed}) now
takes the form
\begin{equation}\label{dyn}
\fl \ddot H H = \frac12 \dot H^2 -3\dot H H^2 + \left( 2k_3 \right)^{-1} \left(
-k_2 H^4+3 m^2 H^2-\rho_{\rm tot} \pm 3 M^3 \sqrt{H^2 -\Lambda_{\rm b}/6}
\right)  \, .
\end{equation}

The two signs in (\ref{dyn}) correspond to two {\it different\/} dynamical
equations. As explained in Sec.~\ref{equations}, the sign is fixed by
specifying one of the two possible ways of embedding the brane in the bulk.

The goal of the present paper is to study the stability of the classical
solutions when vacuum polarization terms are taken into account. The $k_2$-term
in (\ref{rhoq}), which does not contain time derivatives of $H$, can only
change the position of the future stable points. On the contrary, due to the
$k_3$-term in (\ref{rhoq}), some classical solutions can lose stability.
Therefore, for simplicity (and without loss of generality), we set $k_2=0$ in
our calculations. This assumption simplifies the situation significantly since,
in this case, the stationary points of Eq.~(\ref{dyn}) are precisely the
solutions of the classical equation (\ref{main1}) with $\rho = 0$.

For $m=0$ (the Randall--Sundrum brane), Eq.~(\ref{dyn}) becomes
\begin{equation}
\ddot H H = \frac12 \dot H^2 - 3\dot H H^2+ \left. \left( -k_2 H^4 -\rho_{\rm
tot} \pm 3 M^3 \sqrt{H^2 -\Lambda_{\rm b}/6}\right) \right/ 2k_3 \label{RS}
\end{equation}
We can see immediately that the stationary points can exist only for the
``$+$'' sign in (\ref{RS}). This statement is valid also for $k_2 > 0$. A
negative value of $k_2$ leads to a De~Sitter solution, which has no analogs in
the classical case, so we do not consider it here. By linearizing the brane
equation of motion near the classical stationary point, it is easy to find that
the condition for this stationary point to be stable is $k_3<0$, which is the
same condition as in the standard cosmology.

We now turn to the case $m \ne 0$. If the brane has nonzero tension $\sigma$,
the stationary points of (\ref{dyn}) in the case $k_2=0$ can be found by
substituting $\sigma$ into (\ref{main}) and setting $\rho=0$. After that, we
linearize Eq.~(\ref{dyn}) at these stationary points and find the eigenvalues
of the corresponding linearized system. The condition of stability of the
stationary point is that its eigenvalues are negative.

The eigenvalues at the stationary points where $\dot H=0$ are given by
\begin{equation}
\mu_{1,2}=\frac{1}{2} \left( f_1 \pm \sqrt{f_1^2+4f_2} \right) \, , \label{mu}
\end{equation}
where we have made the notation
\begin{equation}
f_1 = -3H \, ,
\end{equation}
\begin{equation}
f_2 = \frac{1}{2k_3} \left(1+\frac{\lambda}{H^2} \pm \frac{2
\ell^{-1}\Lambda_\b/6} {H^2 \sqrt{H^2-\Lambda_\b/6}}\right)\, , \label{f_2}
\end{equation}
\begin{equation}
\lambda = {\sigma \over 3 m^2} \, .
\end{equation}

Two different signs in Eq.~(\ref{f_2}) correspond to two different equations of
motion, while, in Eq.~(\ref {mu}), we have two different eigenvalues of a
single equation.

Since $f_1$ is negative, the eigenvalue $\mu_2$ corresponding to the ``$-$''
sign in Eq.~(\ref{mu}) is also always negative. Moreover, $\mu_1$ is positive
if and only if $f_2$ is positive. As a result, the stability of a fixed point
is equivalent to the condition $f_2 < 0$.

From now on, we consider only the second eigenvalue, $\mu_2$, so the two signs
in the expressions given below always corresponds to two different equations of
motion.

The equations for the fixed points are
\begin{equation}
H-\frac{\lambda}{H} \pm \frac{2\ell^{-1}}{H}\sqrt{H^2-\Lambda_\b/6}=0 \, .
\label{fixed}
\end{equation}

Substituting the value of $\lambda$ obtained from this equation into
(\ref{f_2}), we can rewrite it in the form
\begin{equation}\label{f2}
f_2=\frac{1}{2k_3}\left(2 \pm
\frac{2\ell^{-1}}{\sqrt{H^2-\Lambda_\b/6}}\right).
\end{equation}

We classify solutions corresponding to the upper sign (``$+$'') and the lower
sign (``$-$'') as the ``$+$'' branch and ``$-$'' branch, respectively.  Then,
for the ``$+$'' branch, the expression in the brackets is positive, so we make
our first conclusion:

\begin{itemize}
\item Fixed points of the ``$+$'' branch are stable with respect to quantum
corrections if $k_3<0$.
\end{itemize}

Proceeding further, we solve Eq.~(\ref{fixed}) with respect to $H^2$. The
solutions are given by
\begin{equation} \label{solutions}
H^2=\lambda +2\ell^{-2} \pm 2 \ell^{-1}\sqrt{\lambda-\Lambda_\b/6 + \ell^{-2}},
\label{H^2}
\end{equation}
however, it is necessary to be careful about the two different branches. There
are two possible cases:
\begin{enumerate}
\item $\lambda > \Lambda_\b/6$. In this case, we have one solution for the
``$+$'' branch (with the upper sign in (\ref{solutions})), and one solution for
the ``$-$'' branch (with the lower sign in (\ref{solutions})).

\item $\Lambda_\b/6- \ell^{-2} < \lambda < \Lambda_\b/6$.  In this case, two
solutions (\ref{H^2}) belong to the ``$-$'' branch, while the ``$+$'' branch
does not have fixed points.
\end{enumerate}
For $\lambda < \Lambda_\b/6 - \ell^{-2}$, there are no fixed points for both
equations of motion.

Consider the first situation. We already know that stability of a fixed point
for the ``$+$'' branch requires $k_3<0$. For the ``$-$'' branch, we have
\begin{equation}
f_2=\frac{1}{2k_3}\left(2-\frac{2\ell^{-1}} {\sqrt{\lambda-\Lambda_\b/6 +
2\ell^{-2} + 2\ell^{-1}\sqrt{\lambda - \Lambda_\b / 6 + \ell^{-2}}}}\right)
\label{odin}
\end{equation}
at a fixed point. As we have now $\lambda - \Lambda_\b/6>0$, the second term in
the brackets is always smaller than $\sqrt{2}\,$, so the whole expression in
the brackets is positive. This implies the following conclusion:
\begin{itemize}
\item If the ``$-$'' branch has only one stable point, its stability requires
$k_3<0$.
\end{itemize}

For the second case, both solutions (\ref{H^2}) belong to the ``$-$'' branch.
The upper sign gives the same expression for $f_2$ as in (\ref{odin}). The
condition $\lambda - \Lambda_\b / 6 + \ell^{-2}>0$ makes the second term in the
brackets of (\ref{f2}) smaller then $2$, and again the total number in the
brackets is positive. On the other hand, for the lower sign in (\ref{H^2}), we
have
\begin{equation}
f_2=\frac{1}{2k_3}\left(2-\frac{2 \ell^{-1}}{\sqrt{\lambda - \Lambda_\b / 6 + 2
\ell^{-2} - 2 \ell^{-1} \sqrt{\lambda - \Lambda_\b / 6 + \ell^{-2}}}}\right) \,
. \label{dva}
\end{equation}
One can see that the number in the brackets is negative. As a result, we arrive
at the following conclusion:

\begin{itemize}
\item If two stationary points (\ref{H^2}) belong to the ``$-$'' branch, then
one of them is stable and the other one is unstable.
\end{itemize}

Having in mind the description of the fixed points given in
Sec.~\ref{classical}, we obtain the following picture:

The simplest case corresponds to the classical points described by
Fig.~\ref{neg}b. Note that now the horizontal axis represents $\sigma$, because
we are dealing with the stationary points only. Each equation, with ``$+$'' and
``$-$'' sign, has only one possible stationary point, and these are stable for
$k_3<0$.

For the range of the parameter $\Lambda_{\rm b}$ corresponding to
Fig.~\ref{neg}a, stationary points of the ``$-$'' sign of Eq.~(\ref{dyn}) are
located in the DE part. Their stability also requires $k_3<0$. However,
stationary points of the equations of motion with ``$+$'' sign are located in
the whole admissible region AC. For $\sigma < \rho_c$ (see Fig.~3a), we have
two possible stationary points, one on the AB branch, and the other on the BC
branch. Their stability properties are opposite to each other. In particular,
the unusual condition $k_3>0$ is required for the points on BC branch to be
stable. This conclusion is, however, not so important because the branch BC
cannot be reached as a result of conventional cosmological evolution in the
classical picture.

More interesting situation is realized for $\Lambda_{\rm b}>0$ (see
Fig.~\ref{pos}). Equation (\ref{dyn}) with ``$-$'' sign again has stationary
points on the CD branch, and that with ``$+$'' sign has it on the AC branch. If
$k_3<0$, points on the AB branch are stable, and those on the BC branch are
unstable; for $k_3>0$, the situation is opposite. However, now classical and
quantum dynamics differs significantly. In the classical picture, the BC branch
is part of the evolution track of the Brane\,2, and it can be reached during a
cosmological evolution from some high-energy initial phase. The effects of
vacuum polarization make this branch unreachable for a braneworld with $k_3<0$.

These results can be interpreted from a different point of view. It is
reasonable to assume that the evolution of a brane should be close to the
classical picture in regions which are far removed from singularities. That is
why we are interested only in the case $k_3<0$. What is the fate of a
braneworld if we take quantum corrections into account? In general, this
problem requires numerical integration of Eq.~(\ref{dyn}); however, we can make
several qualitative statements. Fig.~\ref{neg}b contains only stable branches,
so the dynamics with quantum corrections is qualitatively the same, with
recollapse somewhere in the neighbourhood of the points B and D. In the case
shown in Fig.~\ref{neg}a, the Brane\,1 evolution DE does not change much and
results in a recollapse. The Brane\,2 branch AB in the classical picture meets
a quiescent future singularity with $H$ finite and $\dot H \to -\infty$ at the
point B. Such a singularity is absent in the quantum picture because the
expression under the square root in (\ref{dyn}) is positive if $\Lambda_{\rm
b}<0$. Numerical simulations show that the ultimate fate of the Brane\,2
universe in the case with $\Lambda_{\rm b}<0$ and with a large negative
$\sigma$ is also a recollapse.

On the other hand, in the case $\Lambda_{\rm b}>0$, recollapse becomes
impossible since the existence of a square root in (\ref{dyn}) requires
$H^2>\Lambda_{\rm b}/6$. This restriction is valid also for the classical
equation (\ref{main}). However, during the classical evolution, a braneworld
reaches the point C, where $H^2=\Lambda_{\rm b}/6$, with $\dot H=0$, and then
enters the CB branch, which corresponds to super-exponential expansion $({\dot
H > 0})$ but which is impossible in the quantum case. Our numerical results
show that, instead, the universe reaches some point with $H^2= \Lambda_{\rm
b}/6$ and $\dot H<0$. This causes the square root in (\ref{dyn}) to vanish,
which, in turn, leads to the divergence of $\d3h$, while $\ddot H$ remains
finite.\,\footnote{A similar result was obtained in \cite{nojiri1} for the case
of the Randall--Sundrum braneworld model containing matter with unusual
properties resulting in sudden singularity.} As a result, instead of a
quiescent future singularity with $|\dot H| \to \infty$ at the point B, the
universe encounters a much weaker singularity with $\d3h \to \infty$ in the
neighborhood of the point C. Similarly, the Brane\,2 universe misses the point
B and falls into a singularity with $\d3h \to \infty$.

\section{Quiescent singularities in an inhomogeneous
universe: a preliminary analysis} \label{steady}

The preceding discussion focussed on a homogeneous and isotropic universe whose
expansion was governed by the brane equations of motion. Since the real
universe is quite inhomogeneous on spatial scales $\lsim 100$~Mpc, it is
worthwhile to ask whether any of our previous results may be generalized to
this case.

Although we are not yet able to provide a self-consistent treatment of the
brane equations for this important case, still, some aspects of the problem can
be discussed at the phenomenological level. Consider, for instance, the
expansion law (\ref{solution})
\begin{eqnarray} \label{cosmo}
H^2 &=& {\rho + \sigma \over 3 m^2} + {2 \over \ell^2} \left[1 \pm \sqrt{1 +
\ell^2 \left({\rho + \sigma \over 3 m^2} - {\Lambda_{\rm b} \over 6} \right)}
\right] \nonumber \\
&=& {\Lambda_\b \over 6} + {1 \over \ell^2} \left[ \sqrt{ 1 + \ell^2
\left({\rho + \sigma \over 3 m^2} - {\Lambda_\b \over 6} \right) } \pm 1
\right]^2 \, ,
\end{eqnarray}
A necessary condition for the existence of a quiescent singularity is that the
matter density $\rho$ drop to a value which is small enough for the square root
on the right-hand side of (\ref{cosmo}) to vanish. When this happens, the
universe encounters the quiescent singularity at which $\rho$ and $H$ remain
finite, but ${\ddot a}$ and higher derivatives of the scale factor diverge.
Note, however, that, according to (\ref{cosmo}), the universe encounters the
quiescent singularity {\em homogeneously\/}, \ie, every part of the (spatially
infinite) universe encounters the singularity at {\em one and the same\/}
instant of time. This follows from the fact that the density in (\ref{cosmo})
depends only upon the cosmic time and upon nothing else. In practice, however,
the universe is anything but homogeneous, its density varying from place to
place. For instance, it is well known that the density of matter in galaxies is
$\gsim 10^6$ times the average value while, in voids, it drops to only a small
fraction of the average value. This immediately suggests that the brane should
encounter the quiescent singularity in a very inhomogeneous fashion. Underdense
regions (voids) will be the first to encounter the singularity. Even in this
case, since the density in individual voids is inhomogeneously distributed,
more underdense regions lying closer to the void center will be the first to
experience the singularity. It therefore follows that the quiescent singularity
will first form near the centers of very underdense regions. As the void
expands, its density at larger radii will drop below $\rho_s$, where
\begin{equation}
\sqrt{1 + \ell^2 \left({\rho_s + \sigma \over 3 m^2} - {\Lambda_{\rm b} \over
6} \right)} = 0 \, ;
\end{equation}
consequently, the singularity will propagate outward from the void center in
the form of a quasi-spherical singular front. (For simplicity, we have assumed
that all voids have a spherical density profile; this assumption may need to be
modified for more realistic cases; see \cite{sss94,shandarin} and references
therein.)

The above approach provides us with a very different perspective of the
quiescent singularity than that adopted in the previous sections (and in
\cite{SS1}). For one thing, the singularity may be present in certain regions
of the universe {\em right now\/}, so it concerns us directly (as
astrophysicists) and not as some abstract point to which we may (or may not)
evolve in the distant future. The second issue is related to the first, since
the singularity could already exist within several voids (there are as many as
a million voids in the visible universe in at least some of which the condition
$\rho \simeq \rho_s$ could be satisfied), a practical observational strategy
needs to be adopted to search for singularities in voids. (Similar strategies
combined with strenuous observational efforts have led to the discovery of
dozens of black holes in the centers of galaxies \cite{narayan}.)

A number of important issues therefore need to be addressed:

\begin{enumerate}

\item Since $R_{iklm}R^{iklm} \to \infty$ within a finite region at the very
center of a void, it follows that, unless this region is contained within an
event horizon, we will find ourselves staring at a naked singularity\,! (As
shown earlier, quantum effects do soften the singularity so that
$R_{iklm}R^{iklm}$ may remain finite if these effects are included.)

\item In the discussion in Sec.~\ref{quantum}, the issue of particle production
was ignored since it was assumed that we were dealing with conformally coupled
fields which are not created quantum mechanically in the (conformally flat)
homogeneous and isotropic universe which we have been considering up to now.
However, the moment we drop the homogeneity assumption, the issue of particle
production immediately crops up, and we must take it into account if our
treatment is to be at all complete \cite{Zel}.
(In a related context, the quantum creation
of gravitons takes place even in a homogeneous and isotropic universe, since
these fields couple minimally, and not conformally, to gravity \cite{grish75}.)

\end{enumerate}

Let us discuss the possible effect of particle production in more detail.
First, we consider the model of homogeneous universe taking it as an
approximation to the situation inside an underdensity region (void).
Gravitational quantum particle production occurs as the singularity is
approached. Since the local value of the Hubble parameter remains finite at the
singularity, production of the conformally coupled particles (like photons) is
expected to be negligible. However, particles that are non-conformally coupled
to gravity (which could be, for example, Higgs bosons in the Standard Model)
will be copiously produced as the acceleration of the universe $\ddot a$
rapidly increases.   The rate of particle production depends not only on their
coupling to gravity but also on their coupling between themselves.
Gravitationally created primary particles will decay into conformally coupled
secondaries (electrons, photons, neutrino, etc.\@), which will influence the
rate of production of the primaries by causing decoherence in their quantum
state. The whole process is thus not easy to calculate in detail. However, from
very general arguments it can be seen that creation of matter due to quantum
particle production is important for the dynamics of the universe during its
later stages.\,\footnote{Effects of particle production are negligible in the
neighbourhood of the usual cosmological singularity of the Friedmann universe
because the energy density of ordinary matter and radiation strongly diverges
and thus dominates at this singularity \cite{Zel,BD}.  In our case, the energy density of
ordinary matter remains finite during the classical approach to the quiescent
singularity, hence, particle production effects are of crucial significance.}

For the sake of physical simplicity, we restrict ourselves to the case of
vanishing bulk cosmological constant $\Lambda_\b$ and write Eq.~(\ref{cosmo})
in the form
\begin{equation} \label{cosmo1}
H = {1 \over \ell} + \sqrt{\Delta \rho  \over 3 m^2} \, ,
\end{equation}
where
\begin{equation}
\Delta \rho = \rho - \rho_s \, , \quad \rho_s = - \sigma - {3 m^2 \over \ell^2}
\, ,
\end{equation}
and where we have chosen the physically interesting ``$+$'' sign in
Eq.~(\ref{cosmo}). Thus, we have two free parameters in our theory, namely,
$\ell$ and $\rho_s$. The value of $m$ is assumed to be of the order of the
Planck mass. In this case the early-time behaviour of the universe
follows the standard Friedmann model, as can be
seen from (\ref{cosmo}) or (\ref{cosmo1}).

Let the average particle energy density production rate be $\dot \rho_{\rm
prod}$. Then, differentiating Eq.~(\ref{cosmo1}), we obtain
\begin{equation} \label{rate}
\dot H = {\dot {\rho}_{\rm prod} - \gamma H \rho \over 2 \left(3 m^2 \Delta
\rho \right)^{1/2} } \, ,
\end{equation}
where $\gamma > 0$ corresponds to the effective equation of state of matter in
the universe: if $p = w \rho$, then $\gamma = 3 (1 + w)$.  The second term in
the numerator of (\ref{rate}) follows from the conservation law and describes
the effect of the universe expansion on the matter density. (Note that $\rho$
includes contributions from quantum and classical matter.)

In order to qualitatively assess the effects of particle production, let us
examine two fundamentally distinct possibilities.

\begin{enumerate}
\item Suppose that, in the course of evolution, $\Delta \rho \to 0$ is reached
in a finite interval of time.  Since the Hubble parameter is a unique function
of the energy density, given by (\ref{cosmo1}), and since the singularity value
$\rho_s$ of the energy density is approached from above, it follows that $\dot
H \le 0$ in the neighbourhood of the singular point.  In the purely classical
case we find, after setting $\dot \rho_{\rm prod}$ to zero in (\ref{rate}),
that $\dot H \to - \infty$ as the quiescent singularity is approached. It is
well known, however, that particle production effects are sensitive to the
change in the rate of expansion \cite{BD}, and it is expected that $\dot
\rho_{\rm prod}$ will go to infinity as $\dot H \to - \infty$. Since $\dot
{\rho}_{\rm prod} \gg \gamma H \rho$, this will result in $\dot H$ becoming
positive, which contradicts the assumption that $\dot H \le 0$.

Therefore, under the assumption that the critical density $\rho_s$ is reached
in a finite time, the only possibility for $\dot H$ is to {\em remain bounded}.
In other words, the rate of particle production should exactly balance the
decrease in the matter density due to expansion, turning the numerator in
(\ref{rate}) to zero:

\begin{equation} \label{prodrate}
\dot \rho_{\rm prod} - \gamma H \rho \to 0 \quad \Rightarrow \quad \dot
\rho_{\rm prod} \to {\gamma \rho_s \over \ell} \quad \mbox{as} \quad \rho \to
\rho_s \, .
\end{equation}
In this case, the universe reaches its singular state with the energy density
due to particle production
exactly balancing the density decrease caused by expansion, as
given by (\ref{prodrate}).

\item It is not clear whether the above regime will be realized or whether, if
realized, it will be stable, since it requires the exact balancing of rates
(\ref{prodrate}) at the singularity. A second distinct possibility is that, due
to the presence of particle production, the value of $\Delta \rho = \rho -
\rho_s$ always remains bounded from below by a nonzero density. In this
scenario, $\dot H$ initially decreases ($\vert \dot H\vert$ increases) under
the influence of the increasing factor $1/\sqrt{\Delta \rho}$ in (\ref{rate}).
However, a large value of $\vert \dot H\vert$ induces active particle
production from the vacuum which leads to an increase in the value of $\dot
\rho_{\rm prod}$ in (\ref{rate}). As the value of $\Delta \rho$ reaches its
(nonzero) minimum, we have $\dot H = 0$ at this point, according to
(\ref{cosmo1}), after which the rate $\dot H$ becomes positive due to
self-sustained particle production that continues because of the large value of
the second time derivative $\ddot H$. After a period of extensive particle
production, the universe reaches another turning point $\dot H = 0$ after which
is continues to expand according to (\ref{cosmo1}) with decreasing energy
density. Thus, we arrive at a model of cyclic evolution with periods of
extensive particle production alternating with periods of classical expansion
during which quantum particle production is negligible. This scenario bears a
formal resemblance to quasi-steady-state cosmology proposed in a very different
context by Hoyle, Burbidge, and Narlikar \cite{HN}. The particle production
rate in our case is estimated by the quantity $\dot \rho_{\rm prod}$ given in
(\ref{prodrate}), which is approximately the value it takes at the turning
points where $\dot H = 0$. The Hubble parameter in this scenario periodically
varies being of the order of $H \sim \ell^{-1}$, and the energy density is of
the order $\rho_s$, so that particle production rate is
\begin{equation}
\dot \rho_{\rm prod} \sim {\gamma \rho_s \over \ell} \, .
\end{equation}

\end{enumerate}

Our discussion thus far was limited to quantum processes within a single
underdense region (void) which was assumed for simplicity to be perfectly
homogeneous. Let us now (qualitatively) discuss whether this scenario can be
generalized to the real (inhomogeneous) universe. Clearly, the particle
production rate $\dot \rho_{\rm prod}$ in this case should be regarded as being
averaged over the entire universe, to which several significantly underdense
voids are contributing. Equation (\ref{rate}) should therefore be treated as an
ensemble average, where the {\em mean\/} particle production rate depends upon
the distribution as well as dynamics of {\em local\/} underdensity regions. As
a result, equation (\ref{rate}) is not expected to explicitly depend upon the
behaviour of the Hubble parameter and, in principle, particle production can
proceed even in a De~Sitter-like universe, in which the Hubble parameter $H$
remains constant in time. The rate of particle production in this case is given
by equality (\ref{rate}) with zero left-hand side:
\begin{equation}
\dot \rho_{\rm prod} = \gamma H \rho \, .
\end{equation}

The value of the Hubble parameter in such a steady-state universe can be
related to the $\Omega$-parameter in matter
\begin{equation}
\Omega_{\rm m} = {\rho \over \rho_s} = {\rho \over 3 m^2 H^2} \, ,
\end{equation}
where we have used the basic Eq.~(\ref{cosmo1}).  For the average energy density, we
obviously have $\rho - \rho_s \approx \rho$.  Hence,
\begin{equation}
H = \frac 1 \ell + \sqrt{\Delta \rho \over 3 m^2} \approx \frac 1 \ell +
\sqrt{\rho \over 3 m^2} = \frac 1 \ell + H \sqrt{\Omega_{\rm m}} \, ,
\end{equation}
or, finally,\,\footnote{For comparison, the late-time value of the Hubble
parameter in LCDM is \cite{ss00} $H = H_0\sqrt{1-\Omega_{\rm m}}\,$.}
\begin{equation}
H \approx {1 \over \ell \left( 1 - \displaystyle \sqrt{\Omega_{\rm m}} \right)
} \, .
\end{equation}

In principle, one might use these preliminary results to construct a braneworld
version of steady-state cosmology, in which matter is being created at a steady
rate in voids rather than in overdense regions (as hypothesized in the original
version \cite{HN}). This would then add one more model to the steadily growing
list of dark-energy cosmologies \cite{sahni04}. These conclusions must,
however, be substantiated by a more detailed treatment which takes into account
the {\em joint\/} effect of vacuum polarization and particle production near
the quiescent singularity. Clearly, whether one or the other effect dominates
will depend upon the number of conformal and non-conformal fields contributing
to the vacuum, their spin weights, etc. We propose to return to this issue in a
future work.

\section{Conclusions}

Cosmological models based on braneworld gravity have the interesting property
of giving rise to singularities which are not commonly encountered in general
relativity. This is largely due to the possibility of different kinds of
embedding of the brane in the higher dimensional (bulk) space-time. Singular
embedding implies that the expansion (in time) of the brane cannot be continued
indefinitely. The singularities which we have examined in this paper arise
because of this reason. They have the property that, while the density,
pressure, and Hubble parameter on the brane remain finite, higher derivatives
of the Hubble parameter blow up as the singularity is approached. For this
reason, we refer to these singularities as being ``quiescent.'' Despite its
deceptively mild nature, the quiescent singularity is a real curvature
singularity at which the Kretschmann invariant diverges ($R_{iklm}R^{iklm} \to
\infty$). The importance of quantum effects in regions of large space-time
curvature has been demonstrated in a number of papers \cite{BD}, and it should
therefore come as no surprise that these effects can significantly alter the
classical behaviour near the quiescent singularity, as demonstrated by us in
this paper. Examining the vacuum polarization caused by massless conformally
coupled fields, we find that these weaken the quiescent singularity resulting
in an exceedingly mild (soft) singularity at which the values of $H$, ${\dot
H}$, and ${\ddot H}$ remain finite while $\d3h$ and higher derivatives of the
Hubble parameter diverge.

Unlike the classical Big-Bang singularity, the quiescent singularity in
braneworld models is reached in regions of {\em low density\/} and is therefore
encountered during the course of the universe expansion rather than its
collapse. Densities lower than the mean value are known to occupy a large
filling fraction within the cosmic web \cite{shandarin,sheth}.  Therefore, if
the braneworld model is a true representation of reality, one might speculate
that it is likely to encounter the quiescent singularity (or its
quantum-corrected counterpart, the ``soft singularity'') within large
underdense regions, or voids. The rapidly varying space-time geometry near the
quiescent singularity can, in addition to vacuum polarization, also give rise
to quantum creation of fields which do not couple conformally to gravity.
(Examples include massive (leptons, higgs, etc.\@) as well as massless
(gravitons) particles.) A preliminary estimate made by us in this paper shows
that, if this process is sufficiently intensive, then the universe will expand
at late times in a manner which is reminiscent of quasi-steady-state cosmology,
with the Hubble parameter showing oscillations about a constant value. A more
detailed estimate of particle production, however, lies beyond the scope of the
present paper, and we hope to return to it in a future work.

\ack

Petr Tretyakov and Aleksey Toporensky acknowledge RFBR grant 05-02-17450 and
Russian Ministry for Science and Education grant 2338.2003.2. Aleksey
Toporensky also acknowledges support from IUCAA's ``Programme for enhanced
cooperation between the Afro-Asia-Pacific Region.'' Yuri Shtanov and Varun
Sahni acknowledge support from the Indo-Ukrainian program of cooperation in
science and technology sponsored by the Department of Science and Technology of
India and Ministry of Education and Science of Ukraine.

\end{document}